# A gender analysis of top scientists' collaboration behavior: evidence from Italy[1]

Giovanni Abramo

*Laboratory for Studies in Research Evaluation*
*at the Institute for System Analysis and Computer Science (IASI-CNR)*
*National Research Council of Italy*
ADDRESS: Istituto di Analisi dei Sistemi e Informatica, Consiglio Nazionale delle Ricerche, Via dei Taurini 19, 00185 Roma - ITALY
giovanni.abramo@uniroma2.it

Ciriaco Andrea D'Angelo

*University of Rome "Tor Vergata" - Italy and*
*Laboratory for Studies in Research Evaluation (IASI-CNR)*
ADDRESS: Dipartimento di Ingegneria dell'Impresa, Università degli Studi di Roma 'Tor Vergata', Via del Politecnico 1, 00133 Roma - ITALY
dangelo@dii.uniroma2.it

Flavia Di Costa

*Research Value s.r.l.*
ADDRESS: Research Value, Via Michelangelo Tilli 39, 00156 Roma- ITALY
flavia.dicosta@gmail.com

**Abstract**
This work analyzes the differences in collaboration behavior between males and females among a particular type of scholars: top scientists, and as compared to non top scientists. The field of observation consists of the Italian academic system and the co-authorships of scientific publications by 11,145 professors. The results obtained from a cross-sectional analysis covering the five-year period 2006-2010 show that there are no significant differences in the overall propensity to collaborate in the top scientists of the two genders. At the level of single disciplines there are no differences in collaboration behavior, except in the case of: i) international collaborations, for Mathematics and Chemistry - where the propensity for collaboration is greater for males; and ii) extramural domestic collaborations in Physics, in which it is the females that show greater propensity for collaboration. Because international collaboration is positively correlated to research performance, findings can inform science policy aimed at increasing the representation of female top performers.

---

# 1. Introduction

In recent years there has been a worldwide increase in scientific collaborations (Milojevič, 2014). The share of single-authored publications is observed as constantly on the decline (Abt, 2007; Uddin, Hossain, Abbasi, & Rasmussen, 2012), while the average number of authors of a publication has been continuously increasing (Persson, Glänzel, & Danell, 2004; Wuchty, Jones, & Uzzi, 2007; Bukvova, 2010; Gazni, Sugimoto, & Didegah, 2012; Larivière, Gingras, Sugimoto, & Tsou, 2015). Among other motivations, the interaction between scientists of different disciplines and/or organizations is a response to the need to address the complex challenges of science and society (Hall et al., 2018). In addition, the capacity to activate and manage effective collaborations with colleagues, in their own and other institutions, both domestic and international, has become a rewarding factor in the scientist's career development (Petersen, Riccaboni, Stanley, & Pammolli, 2012). Collaboration allows them to participate in broader research projects, gain access to funding, and not least, to improve personal competencies, with positive effects on the quantity and quality of research outputs. It has been shown that in reality, as research collaboration increases, the number of publications (Ductor, 2015; Lee & Bozeman, 2005) and citations (Bidault & Hildebrand, 2014; Li, Liao & Yen, 2013) also increases. Indeed, the link between research collaboration and performance is widely accepted in the literature (He, Geng & Campbell-Hunt, 2009), although fewer studies have tested the aspect of the impact of research performance on the ability to activate collaborations (Abramo, D'Angelo, & Murgia, 2017).

In general, collaboration behavior at the individual level can vary on the basis of contextual factors: first of all with the research discipline concerned (Abramo, D'Angelo, & Murgia, 2013a; Gazni, Sugimoto, & Didegah, 2012; Yoshikane & Kageura, 2004); also with personal factors, such as gender, age, academic rank (Abramo, D'Angelo, & Murgia, 2014; Kyvik & Olsen, 2008; Bozeman & Gaughan, 2011; Gaughan & Bozeman, 2016; Zhang, Bu, Ding & Xu, 2018); and also with social conventions, particularly those concerning the manner of assigning credits and publication authorship (Katz & Martin 1997; Cronin, 2001; Glanzel & Schubert, 2004).

Studies on the effect of gender on scientific collaboration show that women have less extensive collaboration networks than their male counterparts (Cole & Zuckerman 1984; Bozeman & Corley 2004; McDowell, Larry, Singell, & Stater, 2006), van Rijnsoever, Hessels, & Vandeberg, 2008). In addition, there is greater heterogeneity in their individual networks, which on the one hand implies less specialization (Leahey, 2006), and on the other hand favors inter-disciplinary collaboration (Rhoten & Pfirman, 2007; van Rijnsoever et al., 2008). In reality, as Araújo, Araújo, Moreira, Herrmann, and Andrade (2017) show, it is only in the natural sciences that women are more likely than men to have collaborators from other fields.

Some studies indicate that women seem to prefer collaborations with colleagues from other domestic organizations (Moya Anegón et al., 2009), while showing a lower propensity for international collaboration than their male colleagues (Frehill, Vlaicu, & Zippel, 2010; Larivière, Vignola-Gagné, Villeneuve, Gelinas, & Gingras, 2011). Fox, Realff, Rueda, and Morn (2017), surveying women engineers, found that frequency of international research collaboration varies by region, with European women leading the ranking. However, Abramo, D'Angelo and Murgia (2013b) analyzing the scientific production of academics from Italy, found that even if female researchers register a greater capacity to collaborate at intramural and extramural domestic level, there is still a gap with their male colleagues in terms of international collaborations. Addressing similar objectives, Iglič, Doreian, Kronegger and Ferligoj (2017) surveyed Slovenian scientists in four disciplines: mathematics, physics, biotechnology, sociology. Their research shows that while, in general, gender differences in the level of collaboration are not observed, women are probably more connected with colleagues of other research units and departments in the same organization (intramural), while being less connected in terms of international collaborations. These results are in line with the findings of Jadidi, Karimi, Lietz, & Wagner (2017) who analyzed the international community of computer scientists, and by González-Álvarez and Cervera-Crespo (2017), who, investigating scientific production in neuroscience, claimed that the pattern of female collaboration in this field is less international than is the case for male collaboration.

A number of factors have been identified as the main ones responsible for the difference in collaboration behavior between men and women. Among others, the choices of research collaborators is often influenced by mechanisms of gender homophily, which stimulate a search for collaborations primarily among colleagues of the same gender, with whom the individual is more likely to share values and methodological approaches (Boschini & Sjögren, 2007; Ferber & Teiman, 1980; Mcdowell & Smith, 1992). Also, women academics are still a minority in the main disciplines (Hamel, Ingelfinger, Phimister, & Solomon, 2006; Rivellini, Rizzi, & Zaccarin, 2006), and their presence is still less among academics of higher rank (Athanasiou et al., 2016; Gaughan & Bozeman, 2016). Furthermore, the effects of gender discrimination (i.e. under-recognition of women's contributions to science, gender biases in perceptions of publication quality and collaboration interest, gender biases in evaluations of research work) make female scientists less attractive to potential research collaborators (Knobloch-Westerwick, Glynn, & Huge, 2013). The combination of these factors brings about the isolation of female academics, ever more so in departments that are smaller (Mcdowell & Smith, 1992) and have lower percentages of women (Etzkowitz, Kemelgor & Uzzi, 2000). This isolation becomes still more acute given the long history of male overrepresentation in the academic environment (Rhoten & Pfirman, 2007), and could at least partly explain the so-called "productivity gap", a term indicating that male researchers do indeed perform better than women (Fox, 1983; Cole & Zuckerman, 1984; Long, 1987, 1992; Xie & Shauman, 1998, 2004; Mauleón & Bordons, 2006; Lariviere, Ni, Gingras, Cronin & Sugimoto, 2013). In particular, Abramo, D'Angelo, and Caprasecca (2009), showed that the gender productivity gap is noticeable among top scientists (TSs) only.

Keeping in mind the existence of a triangular relationship between gender, collaboration behavior and research performance, in this paper we intend to verify whether gender differences occur on the collaboration behavior of TSs. The current study is part of the stream of works on this matter by the authors' research group, and more specifically a continuation of two previous works. In the first, Abramo, D'Angelo, and Solazzi (2011) showed that TSs are also those who collaborate more abroad, but that the reverse is not always true. In the second, Abramo, D'Angelo and Di Costa (2018) verified whether TSs have a collaboration behavior different from the others. The results from a longitudinal analysis over two successive five-year periods show a strong increase in the propensity to collaborate at domestic level (both extramural and intramural), however this is less for professors who remain or become TS than it is for their lower-performing colleagues. In contrast, the increase in international collaboration behavior is greater for scientists who become or remain top than it is for their peers.

Using the same dataset as this last work, we will extend the analysis to include the gender variable. The objective is to measure collaboration behavior at the "international", "domestic extramural", and "intramural" levels for TSs, to see if this behavior differs by gender from that of their colleagues. The field of observation consists of the Italian academic system and the co-authorships of scientific publications of 11,145 professors over the five-year period 2006-2010, catalogued according to gender, as well as by their scientific field.

The next section further describes the field of observation and the methodology for the study. Section 3 presents the results obtained from the statistical analyses. The paper closes with the conclusions and questions for further examination.

## 2. Data and method

### 2.1 The research performance indicator

A fundamental requirement of this study is the identification of TSs, and therefore the measurement of individual research performance. The citation-based indicator used to measure individual research performance is the Fractional Scientific Strength (FSS). The value of FSS is measured for professors in the sciences of Italian universities for the 2006-2010 period, with citations counted at 30/06/2017. Because the intensity of publication varies across fields, we need to classify the population under observation into research fields. Incidentally, this will allow us also to investigate whether the collaboration behavior of TSs varies across fields. In Italy each professor is classified in one and only one research field named "scientific disciplinary sector" (SDS, 370 in all).[2,3] SDSs are grouped into disciplines named "university disciplinary areas" (UDAs, 14 in all). We define TSs as professors placing among the top 10% by FSS in each SDS.

The FSS is a proxy of the average yearly total impact of an individual's research activity over a period of time. At present we provide the formula to measure FSS, while referring the reader to Abramo and D'Angelo (2014) for a thorough treatment of the underlying microeconomic theory, and all the limits and assumptions embedded in both the definition and the operationalization of the measurement.

$$FSS = \frac{1}{t}\sum_{i=1}^{N}\frac{c_i}{\bar{c}}f_i$$

Where:
$t$ = number of years of work in the period under observation
$N$ = number of publications in the period under observation
$c_i$ = citations received by publication $i$
$\bar{c}$ = average of distribution of citations received for all cited publications in same year and subject category of publication $i$
$f_i$ = fractional contribution of professor to publication $i$.

---

[2] The complete list is accessible on http://attiministeriali.miur.it/UserFiles/115.htm, last accessed 7 May 2019.
[3] In the Italian university system, competitions for recruitment and career advancement are regulated by specific law, and occur at SDS level. Professors are assigned the SDS they competed in.

The fractional contribution equals the inverse of the number of authors in those fields where the practice is to place the authors in simple alphabetical order but assumes different weights in other cases. For the life sciences, widespread practice in Italy is for the authors to indicate the various contributions to the published research by the order of the names in the byline. For the life science SDSs, we then give different weights to each co-author according to their position in the list of authors and the character of the co-authorship (intramural or extramural).[4]

The reader is warned that evaluative scientometrics is based on: i) the axiom that for the production of new knowledge to have an impact "on scientific advancement", it has to be used by other scientists: no use, no impact; and ii) the assumption that citations "certify" the use of prior knowledge. As a consequence, all the usual limits, caveats, and qualifications apply, in particular: i) publications as not representative of all knowledge produced; ii) bibliometric repertories do not cover all publications; and iii) citations are not always certification of real use and representative of all use.

**2.2 Dataset and data source**

The source for data on each professor of Italian universities is the database maintained by the Ministry of Education, Universities and Research (MIUR),[5] which indexes the name, gender, academic rank, field/discipline (SDS/UDA), and institutional affiliation of all professors in Italian universities, recorded at the close of each year.

The bibliographic dataset used to measure FSS is extracted from the Observatory of Public Research (ORP), a database developed by the authors and derived under license from Clarivate Analytics' Web of Science (WoS). Beginning from the raw data of WoS and applying a complex algorithm for disambiguation of the true identity of the authors and reconciliation of their institutional affiliations, each publication is attributed to the Italian university professor that authored it, with a harmonic average of precision and recall (F-measure) equal to 97% (for details see D'Angelo, Giuffrida, & Abramo, 2011).

Because the bibliographic repositories' coverage of research output in arts & humanities and a number of fields within the social sciences is not completely satisfactory (Hicks, 1999; Archambault, Vignola-Gagné, Côté, Larivière, & Gingras, 2006), and particularly so in Italy,[6] our analysis only focuses on the sciences. Professors in the sciences, totaling 39,139, are classified in 9 UDAs, namely 1 - Mathematics and computer science, 2 - Physics, 3 - Chemistry, 4 - Earth sciences, 5 - Biology, 6 - Medicine, 7 - Agricultural and veterinary sciences, 8 - Civil engineering, 9 - Industrial and information engineering.

The dataset used for the analyses, taken directly from Abramo et al. (2018), is a subset of this population, and is made up of professors who satisfy the following two conditions in the period 2001-2010: i) they are permanently on staff over the whole period, at the same university and SDS; and ii) they have authored at least one publication indexed in WoS.[7]

---

[4] It must be noted that different fractional counting across disciplines does not cause any bias, because the top 10% scientists are extracted from each field. To exemplify, if we did not weight the authors' contribution in Cardiology, the top 10% scientists in cardiology might change, but all the remaining TSs (from the other fields) would be exactly the same.

[5] http://cercauniversita.cineca.it/php5/docenti/cerca.php, last accessed 7 May 2019.

[6] It is no surprise that the Italian National Agency for Research Evaluation (ANVUR) does not apply bibliometrics to measure university performance in such disciplines, in the national research assessment excercises (VQR).

Since in UDA 8 the number of female TS professors is too low (only one), we have omitted this UDA. The dataset consists of 11,145 professors (or 28.5% of the total) distributed over 175 SDSs, as indicated in Table 1. Women are just under 30% of the total population, with a peak of 48.9% in Biology and a minimum of 12.9% in Industrial and information engineering. The lower representation of women in the dataset is due in part to the higher incidence of unproductive women in the period under observation. However, women do represent just under 27% of the total dataset (last row, column 4). The comparison between the percentages indicated in columns 4 and 5 indicates a low concentration of females in the restricted group of TSs in almost all UDAs, the sole exception being Physics, in which women represent 13.7% of the total and 12.9% of the TS. The last two columns of Table 1 highlight the different publication intensity across UDAs and, within each UDA, the higher average output of TSs compared to their non-TS colleagues.[8]

*Table 1: Dataset of the analysis, by UDA; in brackets the share of females*

| UDA[†] | No. of SDSs | No. of professors | Dataset | TS | Non-TS | Avg publications per scientist TS | Avg publications per scientist Non-TS |
|---|---|---|---|---|---|---|---|
| 1 | 9 | 3,705 (33.9%) | 1,044 (32.3%) | 127 (12.6%) | 917 (35.0%) | 20.1 | 6.7 |
| 2 | 8 | 2,879 (17.4%) | 1,016 (13.7%) | 93 (12.9%) | 923 (13.8%) | 63.1 | 25.7 |
| 3 | 11 | 3,606 (37.7%) | 1,325 (32.4%) | 156 (17.3%) | 1,169 (34.4%) | 48.8 | 13.5 |
| 4 | 12 | 1,423 (24.5%) | 379 (24.8%) | 43 (18.6%) | 336 (25.6%) | 17.4 | 7.1 |
| 5 | 19 | 5,851 (48.9%) | 1,879 (44.9%) | 233 (24.0%) | 1,646 (47.9%) | 32.9 | 9.2 |
| 6 | 47 | 12,457 (27.3%) | 3,202 (24.5%) | 392 (10.2%) | 2,810 (26.4%) | 47.7 | 10.8 |
| 7 | 29 | 3,545 (31.6%) | 849 (27.7%) | 103 (13.6%) | 746 (29.6%) | 24.3 | 7.8 |
| 9 | 40 | 5,673 (12.9%) | 1,451 (9.6%) | 140 (5.7%) | 1,311 (10.0%) | 40.2 | 11.8 |
| Total | 175 | 39,139 (29.6%) | 11,145 (26.9%) | 1,287 (14.1%) | 9,858 (28.6%) | 39.8 | 11.6 |

[†] *1 - Mathematics and computer science, 2 - Physics, 3 - Chemistry, 4 - Earth sciences, 5 - Biology, 6 - Medicine, 7 - Agricultural and veterinary sciences, 9 - Industrial and information engineering*

## 2.3 The collaboration propensity indicators

In order to assess the collaboration behavior we analyze the nature of co-authorships, adopting the taxonomy described in Abramo, D'Angelo, & Murgia, (2013a). For each academic *i* of the dataset, we measure the propensity to collaborate overall and by type of collaboration, using the following indicators:

- Propensity to collaborate: $C = \frac{cp_i}{N_i}$, where $cp_i$ is the number of publications resulting from collaborations (two or more co-authors in the byline) over the period, and $N_i$ is the total number of publications authored by the academic *i* over the period;

---

[7] We exclude professors who do not publish, because it would make no sense to compare collaboration behavior of those who do not collaborate because they do not publish. It might be questioned whether it makes sense to investigate the collaboration behavior of scientists with one publication only. We do that because, after all, it is the collaboration behavior of that type of scientists. Moreover, they represent only 7.3 percent (6.9 percent among males, and 8.2 percent among females) of the dataset.

[8] For a thorough analysis of the publications per scientist distributions across field, we refer the reader to D'Angelo and Abramo (2015).

- Propensity to collaborate at the intramural level: $CI = \frac{cip_i}{N_i}$, where $cip_i$ is the number of publications resulting from collaborations with other academics belonging to the same university over the period;
- Propensity to collaborate extramurally at the domestic level: $CED = \frac{cedp_i}{N_i}$, where $cedp_i$ is the number of publications resulting from collaborations with scientists belonging to other domestic organizations over the period;
- Propensity to collaborate extramurally at the international level: $CEF = \frac{cefp_i}{N_i}$, where $cefp_i$ is the number of publications resulting from collaborations with scientists belonging to foreign organizations over the period.

These indicators vary between zero (if, in the observed period, the scientist under observation did not produce any publications resulting from the form of collaboration analyzed), and 1 (if the scientist produced all his/her publications through that form of collaboration).[9]

**2.4 Statistical testing**

In order to respond to the research questions, we have used two types of statistical test.

At the aggregate level (overall) we have used the two-sample t-test with unequal variances, to verify if the variation in gender (male vs female) and status (TS vs non-TS) correspond, on average, to variations in the collaboration behavior of scientists. The preliminary *skewness and kurtosis normality tests* have shown that none of the collaboration propensity distributions is normal. This fact does not rise concern, since in large samples the test is valid for any distributions (Lumley, Diehr, Emerson, & Chen, 2002). We have repeated the exercise applying parametric tests which showed exactly the same results.

At UDA level, we have used a non-parametric test (Wilcoxon rank-sum test) because of the varying sizes of UDAs, and in few cases small sizes. Moreover, the Wilcoxon rank-sum test is both valid for data from any distributions, and much less sensitive to outliers than the two-sample t-test (Mann & Whitney, 1947).

**3. Results and analysis**

All the bibliometric measures described above were calculated for the period 2006-2010, for purposes of verifying whether variation in gender (male vs female) and status (TS vs non-TS) correspond to variations in the collaboration behavior of scientists. For this, a t-test was used. The results of the analysis at aggregate level are shown in Table 2, for all types of collaboration considered.

With gender fixed, the differences between the averages (TS vs non-TS) are consistently in favor of non-TS, apart from propensity to collaborate at the international level (CEF) - the latter being the sole exception in which TSs prevail. The female TSs register as follows:

---

[9] Similar indicators are presented by Martín-Sempere, Garzón-Garcia and Rey-Rocha (2008), and Ductor (2015).

- propensity for international collaboration 7.9% higher (30.2% vs 22.3%) compared to female non-TS colleagues;
- propensity for extramural domestic collaboration (although statistically not significant) lower by 2.5% (50.6% vs 53.1%);
- propensity for intramural collaboration lower by 8.8% (71.1% vs 79.9%);
- overall propensity to collaborate lower by 1.4% (97.0% vs 98.4%).

For males, the above four differences show the same signs, respectively at: +7.7%, -1.4%, -7%, -0.5%.

The comparison between women and men shows that there are no significant differences in the propensity to collaborate for the TSs. On the other hand, for the non-TSs, the differences are statistically significant and in favor of women: for their propensity to collaborate in general (98.4% vs 97.4%); for their propensity for extramural domestic collaboration (53.1% vs 50.8%); and for intramural domestic collaboration (79.9% vs 77.0%); the opposite is true for international collaborations (22.3% vs 23.4% in favor of men).

*Table 2: Overall propensity to collaborate relative to status and gender: t-test for comparison of averages (95% confidence interval in brackets).*

|     | F |     |     |     | M |     |     |     |
| --- | --- | --- | --- | --- | --- | --- | --- | --- |
|     | non-TS |     | TS |     | non-TS |     | TS |     |
| C   | 98.4% | [0.980-0.987] | 97.0% | [0.956-0.984] * | 97.4% | [0.971-0.977] | 96.9% | [0.963-0.975] |
| CEF | 22.3% | [0.213-0.234] | 30.2% | [0.266-0.337] *** | 23.4% | [0.227-0.240] | 31.1% | [0.298-0.325] *** |
| CED | 53.1% | [0.518-0.543] | 50.6% | [0.465-0.548] | 50.8% | [0.500-0.516] | 49.4% | [0.478-0.510] |
| CI  | 79.9% | [0.788-0.810] | 71.1% | [0.669-0.752] *** | 77.0% | [0.763-0.777] | 70.0% | [0.683-0.716] *** |

*Statistical significance: *p-value <0.10, **p-value <0.05, ***p-value <0.01*
*C, propensity to collaborate; CEF, propensity to collaborate at international level; CED, propensity to collaborate at extramural domestic level; CI, propensity to collaborate at intramural level*

## 4. Differences among disciplines

The above analysis was repeated at the UDA level, however, given the low number of female TSs in some UDAs (e.g. eight each in Earth sciences and Industrial and information engineering) a non-parametric test was applied: the Mann-Whitney U test. In particular, the *porder* option of the STATA package "Ranksum" command was used. For each indicator of propensity for collaboration, the tables below show the sign of the difference observed between the two sub-sets and the relative statistical significance.

Table 3 shows the analysis for the propensity to collaborate at international level (CEF), in each UDA. With gender fixed (F/M), in the comparison between TS and non-TS, the *porder* option shows positive sign (for both women and men) in all UDAs. In other words, the TSs show a greater propensity to collaborate abroad than their colleagues, regardless of gender or UDA. While for males the test is significant in all UDAs except UDA 2 (Physics), for females it is significant only in UDA 5 (Biology), 6 (Medicine) and 9 (Industrial and information engineering).

However, when status (TS/non-TS) is fixed, the comparison between women and men shows differences varying with the discipline. In particular, among the TSs, women show a lower propensity to collaborate at international level in UDA 1 (Mathematics and computer science) and 3 (Chemistry). In UDAs 2, 7 and 9 the differences are also in favor of men but are not statistically significant; nor are they significant in the other UDAs. On the other hand, considering the non-TS category, women show a CEF that is significantly lower than that of men in UDA 1 (Mathematics and computer science) and 5 (Biology); in the other UDAs the test is not significant.

*Table 2: Differences in the propensity for international collaboration (CEF) by UDA, according to gender and status*

| | F | M | TS | non-TS |
|---|---|---|---|---|
| UDA† | TS vs non-TS | TS vs non-TS | F vs M | F vs M |
| 1 | + | + *** | - ** | - ** |
| 2 | + | + | - | - |
| 3 | + | + *** | - ** | - |
| 4 | + | + ** | + | - |
| 5 | + *** | + *** | + | - *** |
| 6 | + *** | + *** | + | + |
| 7 | + | + *** | - | + |
| 9 | + * | + *** | - | + |

† 1 - Mathematics and computer science, 2 - Physics, 3 - Chemistry, 4 - Earth sciences, 5 - Biology, 6 - Medicine, 7 - Agricultural and veterinary sciences, 9 - Industrial and information engineering
*Statistical significance: *p-value <0.10, **p-value <0.05, ***p-value <0.01*

Table 4 provides the analysis of propensity to collaborate at extramural domestic level (CED). Columns 2 and 3 show that differences in behavior between TSs and non-TSs are only significant in Chemistry (UDA 3) for women, and only in Chemistry and Physics (UDA 2) for men, in both cases in favor of non-TSs. Focusing on TSs, the comparison between women and men shows statistically significant differences in favor of the former only in Physics (UDA 2). On the other hand, analyzing non-TSs, there are significant differences in favor of women in UDA 4 (Earth sciences), 6 (Medicine), 9 (Industrial and information engineering), and in favor of men in Mathematics (UDA 1).

*Table 3: Differences in the propensity to extramural domestic collaboration (CED) by UDA, according to gender and status*

| | F | M | TS | non-TS |
|---|---|---|---|---|
| UDA† | TS vs non-TS | TS vs non-TS | F vs M | F vs M |
| 1 | + | - | + | - ** |
| 2 | + | - *** | + * | - |
| 3 | - ** | - *** | - | - |
| 4 | - | + | + | + ** |
| 5 | - | - | - | - |
| 6 | - | - | - | + * |
| 7 | + | - | + | + |
| 9 | + | + | + | + * |

† 1 - Mathematics and computer science, 2 - Physics, 3 - Chemistry, 4 - Earth sciences, 5 - Biology, 6 - Medicine, 7 - Agricultural and veterinary sciences, 8 - Civil engineering, 9 - Industrial and information engineering
*Statistical significance: *p-value <0.10, **p-value <0.05, ***p-value <0.01*

Finally, Table 5 shows the results for propensity to collaborate at intramural level (CI). In the TS versus non-TS comparison, statistically significant differences were observed for men, and in favor of non-TS, in all the UDAs considered. What emerged at an overall level in the previous section is confirmed at the level of individual disciplines: male TSs show a significantly lower propensity for intramural collaboration than do their non-TS male colleagues. For women, TS versus non-TS comparisons are consistently in favor the latter, but significant in only three UDAs (Chemistry, Biology, Medicine).

The comparison between women and men does not show significant differences for TSs in any of the cases. Instead, concerning non-TSs, in 4 UDAs (1, Mathematics and computer science; 2, Physics; 3, Chemistry; 5, Biology), propensity to collaborate at intramural level is significantly higher for women than for men.

*Table 4: Differences in propensity for intramural collaboration (CI) by UDA, according to gender and status*

|  | F | M | TS | non-TS |
|---|---|---|---|---|
| UDA† | TS vs non-TS | TS vs non-TS | F vs M | F vs M |
| 1 | -      | - *** | + | + *** |
| 2 | -      | - *** | + | + ** |
| 3 | - ***  | - *** | + | + *** |
| 4 | -      | - *** | - | - |
| 5 | - ***  | - *** | + | + *** |
| 6 | - ***  | - *** | - | + |
| 7 | -      | - *** | + | - |
| 9 | -      | - *** | + | - |

† 1 - Mathematics and computer science, 2 - Physics, 3 - Chemistry, 4 - Earth sciences, 5 - Biology, 6 - Medicine, 7 - Agricultural and veterinary sciences, 8 - Civil engineering, 9 - Industrial and information engineering
*Statistical significance: \*p-value <0.10, \*\*p-value <0.05, \*\*\*p-value <0.01*

## 5. Conclusions

Many studies in the literature agree that gender matters, both in the research performance and in the collaboration behavior of scientists. Compared to male colleagues, women seem to prefer collaborations with colleagues from other domestic organizations (both intramurally and extramurally), while they show a lower propensity for international collaborations. It has been shown also that collaboration intensity is positively correlated with research performance and, vice versa, research performance seems a driver of attractiveness for scientific collaborations.

The existence of this triangular relationship between gender, collaboration behavior and research performance, has prompted the authors to check whether gender matters in the collaboration behavior of top performers, as a natural sequel of the authors' previous empirical studies on these interrelated topics.

The test set is composed of 11,145 professors and the coauthorship of their scientific publications over the 2006-2010 period. Examining this data, the average values for propensity to collaborate at domestic level are always lower for TSs than for their non-TS colleagues, both among men and women. On the contrary, the propensity to collaborate internationally sees the TSs prevail, without distinction for gender.

Focusing on the TSs, at the aggregate level the comparison between women and men does not show statistically significant differences in propensity for collaboration, either domestic or international; gender differences do emerge for the non-TS set, for all types of collaboration.

At the level of single disciplines, for TSs, statistically significant gender differences are limited to three cases: in Mathematics and computer science, as in Chemistry, women show a lower propensity to collaborate at the international level; in Physics, men show a lower propensity to collaborate at extramural domestic level. For non-TSs, significant gender differences emerge in some ten cases. Women show less propensity to collaborate at international level in Biology and Mathematics and computer science.

For extramural domestic collaboration the differences are in favor of women in Earth sciences, Medicine, and Industrial and information engineering; in favor of men in Mathematics. Women also show a higher propensity to collaborate at the intramural level in the disciplines of Mathematics and computer science, Physics, Chemistry, and Biology.

For extramural domestic collaboration, the differences are in favor of women in Earth sciences, Medicine, and Industrial and information engineering; in favor of men in Mathematics. Women show a higher propensity to collaborate at intramural level in Mathematics and computer science, Physics, Chemistry, and Biology.

Further to what is known in the literature, the results of the study suggest that the differences in collaboration behavior between males and females do not concern TSs, in particular no differences occur in the propensity to collaborate at the international level. Evidently, the two-way positive link between international collaboration and research performance is confirmed as, differently from female non-TSs, female TSs have a propensity to engage in international collaboration similar to males.

Several gender policies have been envisaged in the Italian research system, as highlighted by the European Institute for Gender Equality (https://eige.europa.eu/gender-mainstreaming/toolkits/gear/legislative-policy-backgrounds/italy). In particular "The National Code of Equal Opportunities between Women and Men", established by Legislative Decree No. 198 in 2006, sets the obligation for Public Administrations (and therefore Universities) to adopt a Positive Action Plan (PAP). The plan lasts three years and must assure the removal of all obstacles hindering equal opportunities at work between men and women. The directive of the Presidency of the Council of Ministers of 23 May 2007 identifies the instruments and the areas of intervention: positive actions aiming at balancing female representation in sectors and professional levels where they are underrepresented; the organisation of work aiming at promoting work-life balance; and hiring and promotional mechanisms targeting women.

Unfortunately, because of everlasting Government instability in Italy, very little (extension of the maternity leave to post-doc researchers) more than declarations of intent has been actually realized.

With regard to the specific focus of this study, few policy mechanisms might be considered. Because, all others equal, increase in productivity is the underlying aim of all productive systems, fostering international collaboration is an indirect way to achieve it. In particular for women, who are noticeably underrepresented among TSs. To foster the propensity of women to collaborate at the international level, a wide variety of incentives can be envisaged. Increasing the freedom and responsibility of individual female researchers to form international research partnerships and attract female foreign researchers. Utilizing honorary and visiting professor or research-fellow appointments to attract female external scholars for collaboration purposes. The creation of internationalization offices, focused on promotion of the institutions research qualities and strengths, with a specific focus on women. Finally, funding schemes can be specifically engineered to require partnerships embedding female individuals, thus facilitating bottom-up collaboration involving women.

In the interpretation of the outcomes of the analysis, we urge scholars to take into account the limitations and assumptions embedded in the bibliometric approach for measurement of research performance and collaboration, and the sensitivity of the results to the conventions and classification schemes adopted, and last but not least the characteristics of the country system under analysis. Given this, the reproduction of this study in other countries would provide interesting interpretive keys on the phenomenon - clearly impacted by the sociocultural features of the different national science systems. Possible future research could investigate the trends of gender differences in TSs' collaboration behavior, through time-series analysis.